\documentclass{aastex631}

\usepackage{comment} 
\usepackage{graphicx} 
\usepackage{color} 
\usepackage{times, ulem}
\usepackage{CJKutf8} 

\def\tabref#1{Table~\ref{#1}}

\def\equref#1{Equation~(\ref{#1})}

\def\figref#1{Figure~\ref{#1}}
\def\Figref#1{Figure~\ref{#1}}
\def\Figsref#1{Figures~\ref{#1}}

\shorttitle{Magnetic field strength of young stars}
\shortauthors{Yamashita, Itoh, and Toriumi (2024)}
\graphicspath{{./}{figures/}}

\begin{document}
\begin{CJK*}{UTF8}{ipxm}

\title{Variations in the magnetic field strength of pre-main-sequence stars, solar-type main-sequence stars, and the Sun}

\author[0009-0004-7156-4943]{Mai Yamashita (山下\ 真依)} %
\affiliation{Interdisciplinary Faculty of Science and Engineering, Shimane University \\
1060, Nishikawatsu, Matue \\
Shimane 690--8504, Japan}
\email{yamashita.mai@riko.shimane-u.ac.jp}

\author[0000-0002-3231-1159]{Yoichi Itoh (伊藤\ 洋一)}
\affiliation{Nishi-Harima Astronomical Observatory, Center for Astronomy, University of Hyogo \\
407--2 Nishigaichi, Sayo-cho \\
Hyogo 679--5313, Japan}


\author[0000-0002-1276-2403]{Shin Toriumi (鳥海\ 森)} 
\affiliation{Institute of Space and Astronautical Science, Japan Aerospace Exploration Agency \\
3--1--1 Yoshinodai, Chuo-ku, Sagamihara \\
Kanagawa 252--5210, Japan}



\begin{abstract}
The surface magnetic fields of pre-main-sequence stars and zero-age main-sequence stars are notably strong, resulting in the generation of numerous spots and the emission of bright chromospheric lines. Rotational variations in magnetic field strength have been identified in T Tauri stars (TTSs) and young main-sequence stars using Zeeman--Doppler imaging. This study investigates the relationship between the mean values and variation amplitudes of the magnetic field strengths of TTSs, main-sequence stars, and the Sun. The findings reveal a positive correlation of over three orders of magnitude, suggesting that a common mechanism drives the magnetic fields of these stars. This positive correlation implies that stars with larger spot sizes experience greater variation amplitudes due to rotational modulations. For the Sun, both the mean magnetic field strength value and its variation amplitude tend to be higher during solar maximum than during solar minimum. 
\end{abstract}



\section{Introduction} \label{sec:intro} 

A sunspot is a region characterized by a strong magnetic field, typically ranging from $2000$ to $4000\ \mathrm{G}$), in contrast to the weak magnetic fields of only a few Gauss found in quiet regions. The dynamo process amplifies the toroidal magnetic field below the stellar surface, with magnetic buoyancy, likely supported by external convection, causing the field to rise. When a magnetic flux tube emerges in the photosphere, its cross-section appears as pairs of sunspots. These flux tubes suppress convection within the spot, preventing the surrounding plasma from entering, which renders sunspots relatively cool and dark, with a temperature of $\approx 4000\ \mathrm{K}$. The strong magnetization of sunspots is confirmed by the Zeeman effect, which is observed in magnetically sensitive absorption lines within these regions \citep{ha08}.

These spots are present not only on the sun but also on stellar surfaces. \citet{h72} proposed the starspot model to explain the wave-like light curves of rotating RS Canum Venaticorum type stars, attributing the variations to dark starspots. Several studies have obtained spectroscopic evidence of starspots such as molecular bands of VO, TiO, and TiO${}_2$ (e.g., \citealt{fa16}) and bright bumps in rotationally broadened absorption lines (e.g., \citealt{m09}; \citealt{matysse}). Photometric evidence has also been obtained; the modulation of the disk-integrated intensity is accompanied by color variations, and the photometric periods are consistent with the rotational velocity. In addition, linear polarization varies in accordance with the photometric period. Strassmeier has provided detailed historical accounts of these studies (1992, 2009). 
The theoretical maximum value of the magnetic strength of pre-main-sequence (PMS) stars was $3000\ \mathrm{G}$ \citep{s94}. In fact, strong magnetic fields of a few thousand $\ \mathrm{G}$ have been observed from PMS stars using Zeeman broadening or Zeeman--Doppler imaging in previous studies (e.g., \citealt{y11}, \citealt{mapp}). 

In recent years, \textit{Kepler} \citep{b10} and the \textit{Transiting Exoplanet Survey Satellite} (\textit{TESS}, \citealt{ri15}) have been used to investigate stellar magnetic activities, such as flares, spots, faculae, activity cycles, rotation, and differential rotation. For example, \citet{m13} detected $365$ flares with \textit{Kepler} data regarding $148$ solar-type stars. The releasing energy of the flare is $100$ times larger than that of solar flares. \citet{n15} found that the Sun and superflare stars show a positive correlation between the amplitude of the light curve and the $r_0$(8542) index, the residual core flux of the Ca II IRT normalized by the continuum level at the line core. This correlation means that superflare stars have large starspots and high magnetic activity compared to the Sun. Moreover, \citet{ik20} implements a computational code for starspot modeling to deduce stellar and spot properties, such as spot emergence and decay rates. 
As an example of the studies for faculae, \citet{r18}, the ratio of the Ca II HK emission line flux to the bolometric flux, $R^{\prime}_{\rm HK}$, show a positive correlation with their variation, $\Delta R^{\prime}_{\rm HK}$ in \citet{b22} and \citet{r18} for the Sun and $72$ Sun-like stars. 
In \citet{na19}, the light variation caused by differential rotation is detected in the \textit{Kepler} light curves. In this study, we discuss spots and activity cycles. 

\citet{r15a} presented \textit{K2} light curves for F-, G-, K-, and M-type zero-age main-sequence (ZAMS) stars in the Pleiades cluster. The amplitudes of the brightness ranged from $0.001\ \mathrm{mag}$ to $0.556 \ \mathrm{mag}$. \citet{y22b} obtained {\it TESS} photometric data of $33$ ZAMS stars in IC 2391 and IC 2602, revealing that the amplitudes of the brightness ranged from $0.001\ \mathrm{mag}$ to $0.145\ \mathrm{mag}$. Both studies indicate that the starspot coverage corresponds to $0.1-21\%$. 
The correlation or anticorrelation between brightness variation caused by starspot and long-term variation (magnetic activity) has also been well studied \citep{m17}. 
The light curves of PMS stars suggest that massive starspots are present on their surfaces. The first study to detect a sinusoidal light curve from a weak line T Tauri star (WTTS), considered to be caused by starspots, was reported by \citet{bo93}. Subsequently, \citet{s03} revealed that another WTTS, V410 Tau, exhibited a large brightness variation of $\Delta V \sim 0.6\  \mathrm{mag}$, suggesting the presence of a massive spot or spot group with a surface coverage of $\sim29\%$ and an average magnetic field strength of $500\ \mathrm{G}$. 
Starspot lifetimes of solar-type main-sequence stars have been observed to range from tens to hundreds of days. The starspot lifetime is measured on solar-type stars from the Kepler data in \citet{ma17}. The lifetime of large spots ranges from $\sim 50 - 300\ \mathrm{days}$, which is longer enough than the rotation period of the star, and shorter than the typical length of observations ($\sim480\ \mathrm{days}$).   In addition, they discovered that the appearance rates of starspots on slowly rotating solar-type stars were not significantly different from those of sunspots. \citet{na19} also investigated 5356 active solar-type stars observed by Kepler and obtained the temporal evolution of 56 individual star spots. The lifetimes of starspots range from $10$ to $350\ \mathrm{days}$. 
Photometry is an effective method for detecting star spots and activity cycles, but it is not a direct method. As the stage of the work presented here, we combine the direct method of magnetic field measurement with the photometric results. 

The activity cycle of the Sun and stars also causes variations in the magnetic field strength for several years (e.g., \citealt{s16}). The Sun is observed to have an 11-year activity cycle. Numerous active regions are observed during the solar maximum, which is defined as the period when the average number of sunspots in the $13\ $month bin is the highest. Active regions may not appear for several months or years during the solar minimum. The Mount Wilson Observatory monitored the strength of the chromospheric emission lines of Ca II in main-sequence and PMS stars and detected variations over several years \citep{r18}. This indirect evidence indicates that stars too undergo magnetic activity cycles. However, measuring the magnetic field strength of the same star every few years is challenging. 

In this study, we investigated the relationship between the mean magnetic field strength and temporal variation of the magnetic field strengths of TTSs, main-sequence stars, and the Sun. The main timescale discussed in this study corresponds to the rotation timescale of the stellar surface. The datasets have observational periods of more than four days, and it is enough to detect the rotational modulation. The typical rotational periods are $\sim7\ \mathrm{days}$, $\sim3\ \mathrm{days}$, and $\sim5\ \mathrm{days}$ for CTTSs, WTTSs, and ZAMS stars with solar mass \citep{ga13}. 
Second, we also discuss the magnetic field variations caused by magnetic activity cycles. The observational period is not long enough to detect the magnetic activity cycles directly. However, each star has a different period of cycles and the phase in the activity cycle is random in each star when it is observed. In the following section, we describe the datasets and the reduction of PMS stars, ZAMS stars, and the Sun. In Section 3, we present the results and discuss the relationship between the mean magnetic field strength and its variation. Finally, Section 4 summarizes the variations in magnetic field strength across PMS stars, main-sequence stars, and the Sun.

\section{Data sets and data reduction}
\subsection{Stars}
We investigated the disk-averaged magnetic field strength $<B>$ (in the line-of-sight direction) and the variation of the magnetic field strength $\Delta B = B_{\rm max} - B_{\rm min}$. 
Measurements of stellar magnetic fields are usually challenging. Currently, two groups have succeeded in providing systematic studies. One group conducted polarization spectroscopy and Zeeman--Doppler imaging with ESPaDOnS using Canada’s Hawaii Telescope (CFHT), SPIRou on CFHT, and NARVAL on Telescope Bernard Lyot (e.g., \citealt{f16}, MaPP project, MaTYSSE project). Another study conducted high-resolution spectroscopy and detected Zeeman broadening with a high-resolution spectrometer such as the Sandiford cross-dispersed echelle spectrometer on the $2.1\ \mathrm{m}$ Otto Struve Telescope, CSHELL spectrometer on the NASA Infrared Telescope Facility, and the Phoenix high-resolution near-IR spectrometer on the $8.1\ \mathrm{m}$ Gemini South Telescope (e.g., \citealt{y11}). Moreover, the magnetic field strength varies with the rotational phase. 

\subsubsection{PMS stars}

The magnetic field strength for thousands of stars have already been measured. However, the number of objects whose magnetic field variation has been measured is limited. Such observations are few and rare. 
Furthermore, it becomes more rare when restricted to PMS stars. In this study, we referred the MaPP project \citep{mapp}, MaTYSSE project (\citealt{matysse}, \citealt{yu17}), \citet{f16}, and \citet{v18}, which systematically examined the mean magnetic strength, $<B>$, and the magnetic field strength, $\Delta B$ of PMS stars. All the data were taken from the referred papers which were based on the Zeeman-Doppler imaging observations with CFHT/EsPaDOnS and its direct copy, TBL/Narval. We consider that the measured $<B>$ and $\Delta B$ have sufficiently high uniformity. Most of the referred papers conducted more than seven observations. \citet{v18} had a maximum of 16 observations, and some objects had only two observations. During the observations, they used the same instruments for each star. 

The targets are $28\ $PMS stars, which belong to seven molecular clouds (Tau, Lup, IC 1, Ori, Sco, and Upper Sco) and five moving groups (NGC 6530, R Vul R2, TWA, ${\rm \beta}$ Pic, and Columba), except for V4046 Sgr A and B. 
Their masses and ages were in the range $\sim0.5 - 3.8\ \mathrm{M_\odot}$ and $\sim10^5 - 10^7\ \mathrm{yr}$, respectively (referred from the cited papers). The typical error of the masses is $\pm0.05\ \mathrm{M_\odot}$ in \citet{f16}, $\pm0.1\ \mathrm{M_\odot}$ in \citet{v18}. 

The timescale of observations for CTTS, WTTS, TTS, and PTTS ranges $4\ \mathrm{days}$ to $2\ \mathrm{years}$, except for $6\ \mathrm{years}$ of V4046 Sgr A and B. We did not include the objects with non-significant variation in magnetic field strength. The uncertainties of $\Delta B$ are not described in the previous studies, however, the uncertainties of $\Delta B$ are considered to be several Gauss, because the uncertainties of $B_l$, longitudinal magnetic field strength, is about several Gauss according to Figure A2 of \citet{f16}. Their observations covered all the rotational phases, and systematic errors leading to underestimation are considered small. In Figure 2 of \cite{d10}, one of the MaPP project and AA Tau was observed, the error in $B_l$ range about a few dozen gausses to $\sim 300\ \mathrm{G}$. AA Tau has $<B>$ of $1300\ \mathrm{G}$, then their variation is considered to be significant.

\subsubsection{Main-sequence stars}

We referred the systematic study of $<B>$ and $\Delta B$ for the ZAMS, young main-sequence stars, and main-sequence stars with long-term (multi-year) magnetic variability \citep{f16, f18}. Their properties are described in \tabref{tab:ysoB}. The measurements of $<B>$ and $\Delta B$ have sufficiently high uniformity because all the referred papers conducted Zeeman-Doppler imaging with CFHT/EsPaDOnS and TBL/Narval. Most of the referred papers conducted more than seven observations. 

The timescale of observations for ZAMS, and young main-sequence stars are $4\ \mathrm{days}$ to half year. That of main-sequence stars with long-term (multi-year) magnetic variability are half a year to $9\ \mathrm{years}$. The uncertainties of $\Delta B$ are considered to be several Gauss, because the uncertainties of each measured magnetic field strength are about several Gauss according to Fig. A2. of \citet{f16} and \citet{f18}. Their observations covered all the rotational phases, and systematic errors leading to underestimation are considered small. 

The targets are $42\ $ main-sequence stars. The ZAMS stars and young main-sequence stars belong to the moving group AB Dor and the four open clusters Pleiades, Her-Lyr, Coma Ber, and Hyades. The details are described in \citet{f16} and \citet{f18}. They selected the targets from a list of members in nearby stellar associations or clusters, and only stars with published rotation periods. The most of main-sequence stars with long-term (multi-year) magnetic variability do not belong to any cluster or group. Their masses were in the range $\sim0.7 - 1.6\ \mathrm{M_\odot}$ \citep{f16, f18}. The typical error of the masses is $\pm0.05\ \mathrm{M_\odot}$.

\subsubsection{Stellar properties}

\begin{figure}[htbp]
		\centering
		\includegraphics[width=12cm]{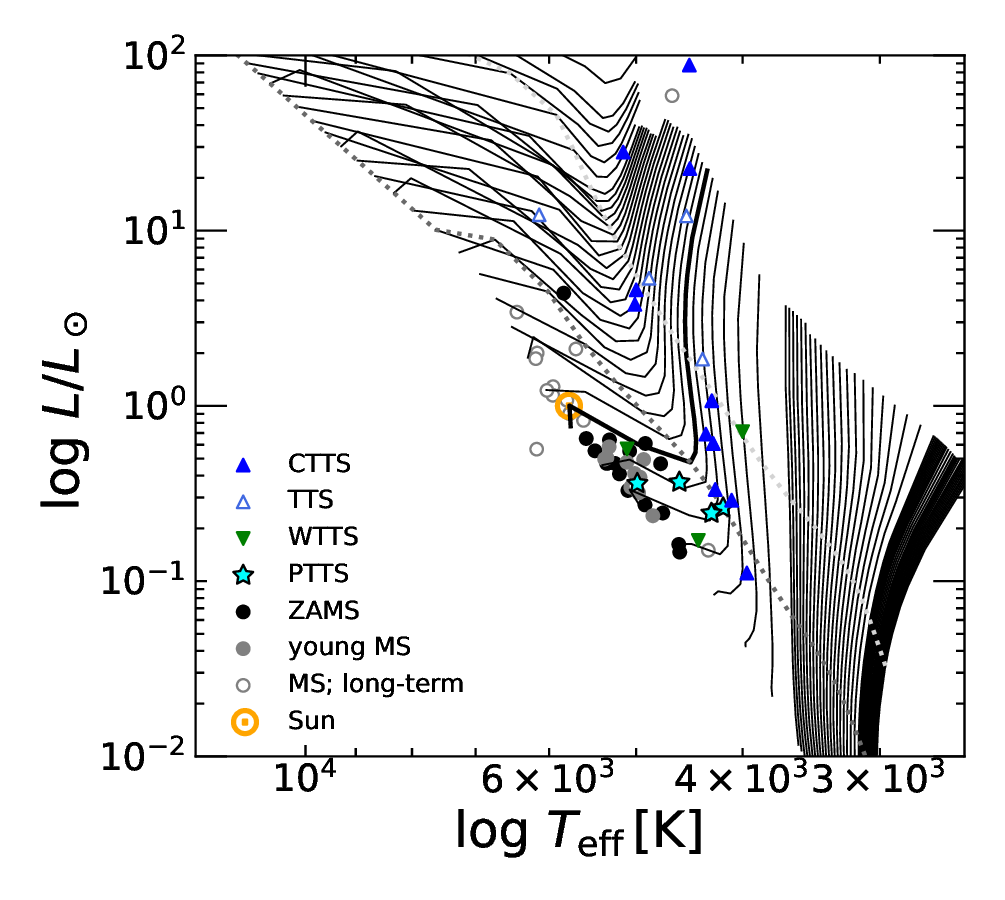} 
		\caption{HR diagram of the investigated PMS, ZAMS, and main-sequence stars. The solid and dot lines denote the evolutionary tracks and isochrones of $10\ \mathrm{Myr}$ and $100\ \mathrm{Myr}$ of \citet{jk07}, respectively. The blue triangles, white triangles, green inverted triangles, and cyan star symbols represent CTTSs, TTSs, WTTSs, and PTTSs, respectively. The filled circles, gray circles, open circles, and circled dots show ZAMS stars, young main-sequence (MS) stars, and main-sequence stars with long-term (multiyear) magnetic variability, and the Sun, respectively. }\label{fig:hrdb}
		\end{figure}

\Figref{fig:hrdb} presents the Hertzsprung--Russell (HR) diagram of the objects investigated. The stellar effective temperature and luminosity are referred from Gaia DR2 \citep{g18}. We note that the luminosity provided by Gaia DR2 (column (6) in \tabref{tab:ysoB}) has already been corrected for the line-of-sight extinction in G-band, $A_G$, which is inferred based on the stellar parallaxes and galactic extinction maps. On the other hand, some previous studies measured interstellar extinction at $V-$band, $A_{V}$ for each star (\citealp{he14}, \citealp{go18}, \citealp{yu23}). It is considered to reflect the small structure of dust and the intrinsic component due to molecular clouds and protoplanetary disks. Then we canceled the Gaia correction ($A_G$) and recorrected with $A_{V}$ for the objects for which $A_{V}$ was measured: 
\begin{equation}
\mbox{new} L = 10^{\frac{A_{V} - A_G }{2.5}} L.
\end{equation}
In other words, $L = \mbox{new} L$ if $A_{V}$ is not measured or if both $A_{V}$ and $A_G$ are not measured. We checked binaries or triplets listed in Gaia DR3 \citep{g23}, and listed in \tabref{tab:ysoB}. 

\begin{longtable}[tbhp]{llccccccccccc}
\caption{\label{tab:ysoB} Mean magnetic strength, $<B>$, and their variation, $\Delta B $, and physical parameters of PMS and ZAMS stars, and main-sequence stars.}\\
\hline\hline
Objects  & Group & $<B>$ & $\Delta B$ & $T_{\rm eff}$ & $L$ & $A_{V}$ & $A_{G}$ & new $L$ & Age & Mass & Binary & $P_{\rm cycle}$\\
{}  & {} & $\mathrm{[G]}$ & $\mathrm{[G]}$ & $\mathrm{[K]}$ & $\mathrm{[L_\odot]}$ & $\mathrm{[mag]}$ & $\mathrm{[mag]}$ & $\mathrm{[L_\odot]}$ & $\mathrm{[Myr]}$ & $\mathrm{[M_\odot]}$ & {} & $\mathrm{[yr]}$\\
(1) & (2) & (3) & (4) & (5) & (6) & (7) & (8) & (9) & (10) & (11) & (12) & (13) \\
\hline
\endfirsthead
\caption{continued.}\\
\hline 
Objects  & Group & $<B>$ & $\Delta B$ & $T_{\rm eff}$ & $L$ & $A_{V}$ & $A_{G}$ & new $L$ & Age & Mass & Binary & $P_{\rm cycle}$\\
{}  & {} & $\mathrm{[G]}$ & $\mathrm{[G]}$ & $\mathrm{[K]}$ & $\mathrm{[L_\odot]}$ & $\mathrm{[mag]}$ & $\mathrm{[mag]}$ & $\mathrm{[L_\odot]}$ & $\mathrm{[Myr]}$ & $\mathrm{[M_\odot]}$ & {} & $\mathrm{[yr]}$\\
(1) & (2) & (3) & (4) & (5) & (6) & (7) & (8) & (9) & (10) & (11) & (12) & (13) \\
\hline
\endhead
\hline
\endfoot
\hline \hline
\multicolumn{2}{@{}l@{}}{\hbox to0pt{\parbox{176mm}{\small References for parameters: (3)(4)(11) Mean magnetic field strength, variation in magnetic field strength, and Mass: MaPP project \citep{mapp}, MaTYSSE project \citep{matysse}, \citet{f16}, \citet{f18}, and \citet{v18}. (5)(6)(8) Effective temperature, luminosity, and line-of-sight extinction in G-band: Gaia DR2 \citep{g18}. (7) Interstellar extinction at $V-$band: \citet{he14}, \citet{go18}, \citet{yu23}. (10) Age: \citet{f16} and \citet{f18}for PTTS, ZAMS, and young main-sequence stars. \citet{b07}, \citet{b15}, \citet{f10}, Gaia DR3 \citep{g23}, \citet{g00}, \citet{ro16}, and \citet{s00} for main-sequence stars with long-term (multiyear) magnetic variability. (12) Binary flag: Gaia DR3 \citep{g23}. ``s'' means single star, and ``b'' means binary or multiple star. (13) The period of the activity cycle referred in the previous studies. ``x'' means an object not detected activity cycles with the long-term observation. ``-'' means an object that was not observed for a long-term. }\hss}} 
\endlastfoot
CTTS                    &            &                            &     &      &       &       &      &       &      &     &   &                 \\ \hline
AA Tau                  & Tau        & 1300                       & 310 & 4267 & 0.74  & 0.40  &      & 1.07  & 1.3  & 0.5 & s & 9.6             \\
BP Tau                  & Tau        & 1300                       & 630 & 4320 & 0.46  & 0.45  &      & 0.69  & 3.6  & 0.9 & s & x               \\
DN Tau                  & Tau        & 530                        & 150 & 3964 & 0.53  & 0.55  & 2.25 & 0.11  & 19.0 & 0.5 & s & 4.1             \\
GQ Lup                  & Lup        & 1300                       & 525 & 4093 & 0.85  & 1.60  & 2.76 & 0.29  & 6.4  & 0.5 & s & -               \\
TW Hya                  & TWA        & 1500                       & 350 & 4236 & 0.26  & 0.27  &      & 0.33  & 9.4  & 0.9 & s & -               \\
V4046 SgrA              & (isolated) & 230                        & 100 & 4254 & 0.61  & 0.00  &      & 0.61  & 3.0  & 0.7 & s & x               \\
V4046 SgrB              & (isolated) & 180                        & 80  & 4254 & 0.61  & 0.00  &      & 0.61  & 3.0  & 0.7 & s & x               \\
CO Ori                  & IC 1       & 84                         & 24  & 4468 & 3.58  & 2.00  &      & 22.59 & 0.04 & 3.2 & s & x               \\
GW Ori                  & Ori        & 66                         & 33  & 4473 & 26.56 & 1.30  &      & 87.96 & 0.02 & 3.8 & s & 0.2             \\
RY Ori                  & Ori        & 53                         & 106 & 5000 & 3.09  & 0.43  &      & 4.59  & 1.4  & 1.7 & s & -               \\
V 1044 Ori              & Ori        & 42                         & 35  & 5138 & 4.45  & 2.00  &      & 28.06 & 0.2  & 2.1 & s & -               \\
V 1149 Sco              & Sco        & 20                         & 15  & 5013 & 2.51  & 0.45  &      & 3.79  & 1.8  & 1.6 & s & x               \\ \hline

WTTS                    &            &                            &     &      &       &       &      &       &      &     &   &                 \\ \hline
TAP 26                  & Tau        & 75                         & 300 & 4389 & 0.28  & 0.25  & 0.79 & 0.17  & 40.0 & 1.0 & s & x               \\
HD 137059               & Lup        & 37                         & 9   & 5096 & 0.57  &       &      & 0.57  & 25.0 & 1.9 & s & -               \\
V 1000 Sco              & Upper Sco  & 51                         & 175 & 4000 & 0.71  &       &      & 0.71  & 1.0  & 1.6 & b & x               \\ \hline

TTS                     &            &                            &     &      &       &       &      &       &      &     &   &                 \\ \hline
BN Ori                  & Ori B      & 201                        & 403 & 6127 & 12.31 &       &      & 12.31 & 3.2  & 3.2 & s & x               \\ 
Cl* NGC 6530 SCB 7      & NGC 6530   & 11                         & 5   & 4866 & 5.35  &       &      & 5.35  & 0.7  & 1.2 & s & -               \\
PX Vul                  & R Vul R2   & 116                        & 8   & 4500 & 12.13 &       &      & 12.13 & 0.1  & 2.5 & s & -               \\
V 1002 Sco              & Upper Sco  & 23                         & 47  & 4350 & 1.29  & 0.39  &      & 1.85  & 0.7  & 1.7 & s & x               \\ \hline

PTTS                    &            &                            &     &      &       &       &      &       &      &     &   &                 \\ \hline
HIP 12545               & beta Pic   & 115                        & 40  & 4167 & 0.26  &       & 1.10 & 0.26  & 24   & 1.0 & s & 3.4             \\
TYC 6349-0200-1         & beta Pic   & 59                         & 95  & 4270 & 0.24  &       &      & 0.24  & 24   & 0.9 & s & 3.2             \\
TYC 6878-0195-1         & beta Pic   & 55                         & 92  & 4567 & 0.37  &       & 1.04 & 0.37  & 24   & 1.2 & s & -               \\
BD-16351                & Columba    & 49                         & 116 & 4988 & 0.61  & 0.08  & 0.65 & 0.36  & 42   & 0.9 & s & 1.7             \\ \hline

ZAMS star               &            &                            &     &      &       &       &      &       &      &     &   &                 \\  \hline
LO Peg                  & AB Dor     & 139                        & 150 & 4564 & 0.16  & 0.09  & 0.17 & 0.15  & 120  & 0.8 & s & 4.8             \\
PW And                  & AB Dor     & 125                        & 175 & 4908 & 0.27  & 0.09  & 0.08 & 0.27  & 120  & 0.9 & s & 8               \\
HIP 76768               & AB Dor     & 112                        & 66  & 4573 & 0.16  &       &      & 0.16  & 120  & 0.8 & s & -               \\
TYC 0486-4943-1         & AB Dor     & 25                         & 48  & 4729 & 0.21  & 0.29  & 0.11 & 0.24  & 120  & 0.8 & s & -               \\
TYC 5164-567-1          & AB Dor     & 63                         & 46  & 5229 & 0.43  & 0.25  & 0.16 & 0.47  & 120  & 0.9 & s & -               \\
BD-07 2388              & AB Dor     & 195                        & 440 & 4905 & 0.61  &       & 0.64 & 0.61  & 120  & 0.9 & s & -               \\
HIP10272                & AB Dor     & 21                         & 28  & 5288 & 0.64  &       & 0.13 & 0.64  & 120  & 0.9 & s & -               \\
HD 6569                 & AB Dor     & 25                         & 19  & 5087 & 0.34  & 0.11  & 0.16 & 0.33  & 120  & 0.9 & s & -               \\
HII 739                 & Pleiades   & 15                         & 28  & 5820 & 4.40  &       &      & 4.40  & 125  & 1.2 & s & -               \\
Cl* Melotte 22 PELS 031 & Pleiades   & 44                         & 37  & 4750 & 0.57  & 0.07  & 0.29 & 0.47  & 125  & 1.0 & s & -               \\
HII 296                 & Pleiades   & 80                         & 50  & 5069 & 0.48  & 0.26  & 0.11 & 0.55  & 125  & 0.9 & s & -               \\
V447 Lac                & Her-Lyr    & 39                         & 14  & 5319 & 0.47  & 0.07  & 0.06 & 0.47  & 257  & 0.9 & s & -               \\
DX Leo                  & Her-Lyr    & 29                         & 36  & 5300 & 0.47  & 0.04  & 0.04 & 0.47  & 257  & 0.9 & s & 4.1             \\
V439 And                & Her-Lyr    & 13                         & 11  & 5552 & 0.65  &       & 0.03 & 0.65  & 257  & 1.0 & s & -               \\
HH Leo                  & Her-Lyr    & 28                         & 33  & 5451 & 0.58  & 0.05  & 0.09 & 0.55  & 257  & 1.0 & s & -               \\
EP Eri                  & Her-Lyr    & 34                         & 11  & 5178 & 0.41  &       & 0.11 & 0.41  & 257  & 0.9 & s & -               \\ \hline

\multicolumn{2}{l}{Young main-sequence star} &                            &     &      &       &       &      &       &      &     &   &                 \\ \hline
Cl* Melotte 111 AV 1693 & Coma Ber   & 33                         & 20  & 5341 & 0.50  & 0.03  & 0.03 & 0.50  & 584  & 0.9 & s & -               \\
Cl* Melotte 111 AV 1826 & Coma Ber   & 25                         & 25  & 5056 & 0.36  & 0.04  & 0.08 & 0.34  & 584  & 0.9 & s & -               \\
Cl* Melotte 111 AV 2177 & Coma Ber   & 10                         & 15  & 5099 & 0.50  & 0.03  & 0.07 & 0.48  & 257  & 0.9 & b & -               \\
Cl* Melotte 111 AV 523  & Coma Ber   & 22                         & 12  & 4828 & 0.23  & 0.04  & 0.02 & 0.24  & 584  & 0.8 & s & -               \\
TYC 1987-509-1          & Coma Ber   & 25                         & 16  & 5316 & 0.50  & 0.07  & 0.06 & 0.51  & 584  & 0.9 & s & -               \\
Mel25 151               & Hyades     & 23                         & 23  & 4922 & 0.49  &       & 0.60 & 0.49  & 584  & 0.9 & s & x               \\
Mel25 179               & Hyades     & 26                         & 26  & 4974 & 0.39  & 0.17  & 0.24 & 0.39  & 625  & 0.9 & s & x               \\
Mel25 21                & Hyades     & 12                         & 18  & 5292 & 0.55  & 0.19  & 0.13 & 0.58  & 625  & 0.9 & s & -               \\
Mel25 43                & Hyades     & 8                          & 14  & 4960 & 0.42  & -0.10 & 0.13 & 0.32  & 625  & 0.9 & b & -               \\
Mel25 5                 & Hyades     & 13                         & 15  & 5016 & 0.40  & 0.16  & 0.12 & 0.41  & 625  & 0.9 & s & x               \\ \hline

\multicolumn{6}{l}{Main-sequence stars with long-term (multi-year) magnetic variability}   &       &      &       &      &     &   &                 \\ \hline
61 Cyg A                & -          & 8                          & 9   & 4298 & 0.15  &       &      & 0.15  & 2000 & 0.7 & s & 7               \\
EK Dra                  & Pleiades   & 75                         & 38  & 5584 & 0.90  & 0.05  & 0.14 & 0.83  & 100  & 1.0 & s & 9.2             \\
HD 179949               & -          & 2                          & 0   & 6153 & 2.00  &       & 0.17 & 2.00  & 4827 & 1.2 & s & x               \\
HD 189733               & -          & 31                         & 24  & 5015 & 0.35  &       & 0.16 & 0.35  & 1600 & 0.8 & s & x               \\
HD 190771               & -          & 7                          & 8   & 5782 & 1.08  &       & 0.11 & 1.08  & 9189 & 1.0 & s & x               \\
HD 35296                & -          & 16                         & 6   & 6171 & 1.86  &       & 0.16 & 1.86  & 40   & 1.2 & s & $> 25$ \\
HD 78366                & -          & 5                          & 5   & 5952 & 1.29  &       & 0.08 & 1.29  & 5212 & 1.1 & s & 12.6            \\
HN Peg                  & Her-Lyr    & 17                         & 13  & 5955 & 1.15  &       & 0.06 & 1.15  & 250  & 1.1 & s & 6.2             \\
II Peg                  & -          & 364                        & 148 & 4600 & 2.06  &       &      & 2.06  & -    & 0.9 & s & 10              \\
NZ Lup                  & -          & 36                         & 18  & 5673 & 2.11  &       & 0.53 & 2.11  & 17   & 1.3 & s & -               \\
$\chi^1$ Ori & Ursa major & 16                         & 7   & 6028 & 1.23  &       & 0.06 & 1.23  & 300  & 1.1 & s & $> 25$ \\
$\eta$ Eri & -          & 14                         & 10  & 4637 & 58.85 &       & 0.17 & 58.85 & 440  & 1.6 & s & -               \\
$\kappa^1$ Cet & -          & 23                         & 5   & 5749 & 0.90  &       & 0.10 & 0.90  & 650  & 1.0 & s & 5.9             \\
$\xi$ Boo A & -          & 13                         & 10  & 6160 & 0.57  &       & 0.11 & 0.57  & 4755 & 1.0 & s & 17.2            \\
$\tau$ Boo & -          & 2                          & 2   & 6420 & 3.43  &       &      & 3.43  & 1000 & 1.3 & s & 0.32           \\
\end{longtable}

\subsection{The Sun -- SOHO and SDO data}

\begin{figure}[htbp]
        	  \centering
            \includegraphics[width=8cm]{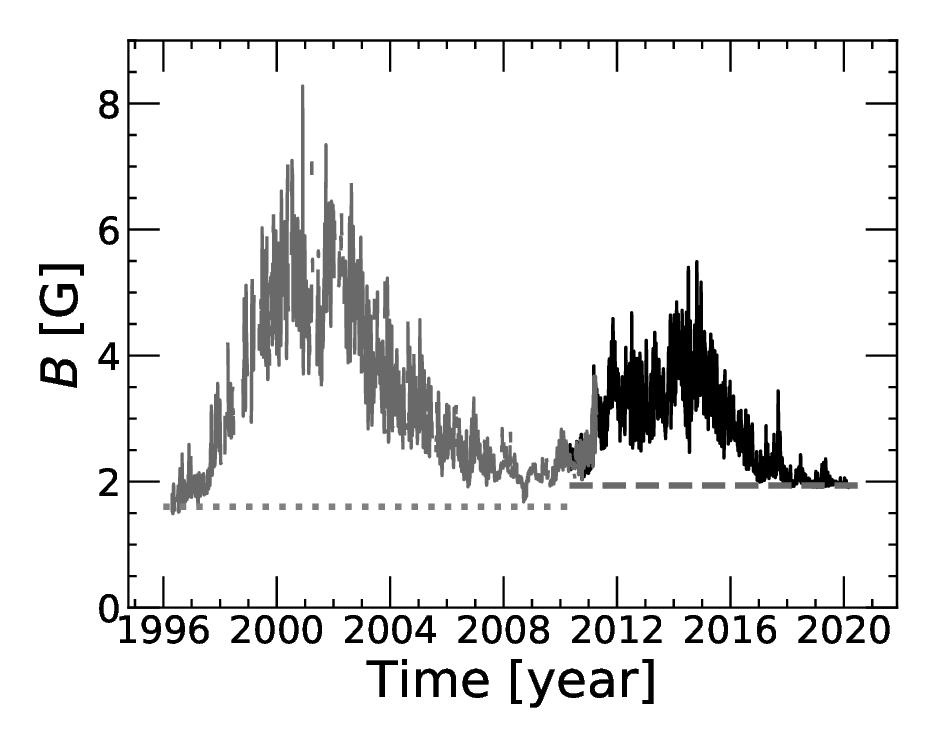}
            \caption{ Time series of averaged unsigned magnetic strength of the Sun. The gray solid and dotted lines denote the data obtained by SDO/HMI and the basal flux, respectively. The black solid and dashed lines denote the data obtained by SOHO/MDI and the basal flux, respectively.  } \label{sdo}
        \end{figure}

We analyzed the time series of the averaged unsigned magnetic strength of the Sun from 1996 to 2019 (\figref{sdo}; \citealt{ta22}, \citealt{ta22b}, and \citealt{na23}). The average unsigned magnetic strength from April 1996 to May 2010 was obtained using the Michelson Doppler Imager (MDI) onboard the \textit{Solar and Heliospheric Observatory (SOHO)}, and that from May 2010 to February 2020 was obtained using the Helioseismic and Magnetic Imager (HMI) onboard the \textit{Solar Dynamics Observatory (SDO)}. The data from the two satellites cover two solar cycles: solar cycle 23 from August 1996 to December 2008 and solar cycle 24 from December 2008 to December 2019. In solar cycles 23 and 24, the solar maxima are at Sep. 2001 and Feb. 2014. However, the periods with the highest number of sunspots without averaging were November 2001 and April 2014. 

The basal flux level of approximately $\sim 2\ \mathrm{G}$ was subtracted from the magnetic field strength (see \citealt{ta22} for further details). This basal flux is assumed to reflect the quiescent background magnetic field of the Sun. For comparing solar data with stellar data, the variation in the magnetic field strength was evaluated. Given the solar rotation period of approximately $27\ $ d, the data were divided into yearly intervals, encompassing multiple rotational periods. The peak-to-peak amplitude of the magnetic field variability $\Delta B \ \mathrm{[G]}$ was calculated by subtracting the 90th and 10th percentiles. The results are presented in Table \tabref{tab:sunB}. 

\begin{table}[htbp]
\centering
\caption{ Mean magnetic strength, $<B>$, and their variation, $\Delta B $, of the Sun in approximately a year bin. }
\begin{tabular}{ccc} 
\hline
start date  & $<B> - {\rm(basal\ flux)} \ \mathrm{[G]}$ & $\Delta B \ \mathrm{[G]}$ \\ \hline
1996-01-01 & 0.214& 0.583 \\
1996-12-19 & 0.536& 0.906 \\
1997-12-07 & 1.575& 1.299 \\
1998-11-25 & 2.901& 1.660 \\
1999-11-13 & 3.627& 1.830 \\
2000-10-31 & 3.276& 1.872 \\
2001-10-19 & 3.712& 1.936 \\
2002-10-07 & 2.562& 1.831 \\
2003-09-25 & 1.889& 1.408 \\
2004-09-12 & 1.510& 1.008 \\
2005-08-31 & 1.004& 0.750 \\
2006-08-19 & 0.784& 0.620 \\
2007-08-07 & 0.611& 0.371 \\
2008-07-25 & 0.467& 0.442 \\
2009-07-13 & 0.632& 0.574 \\
2010-07-01 & 0.691& 1.027 \\
2011-06-19 & 1.400& 1.006 \\
2012-06-06 & 1.394& 1.275 \\
2013-05-25 & 1.704& 1.281 \\
2014-05-13 & 1.808& 1.520 \\
2015-05-01 & 1.171& 1.070 \\
2016-04-18 & 0.480& 0.754 \\
2017-04-06 & 0.273& 0.585 \\
2018-03-25 & 0.085& 0.198 \\
2019-03-13 & 0.060& 0.151 \\ \hline
\end{tabular}
\label{tab:sunB}   
\end{table}

\section{Results and Discussion}
\begin{figure*}
\gridline{\fig{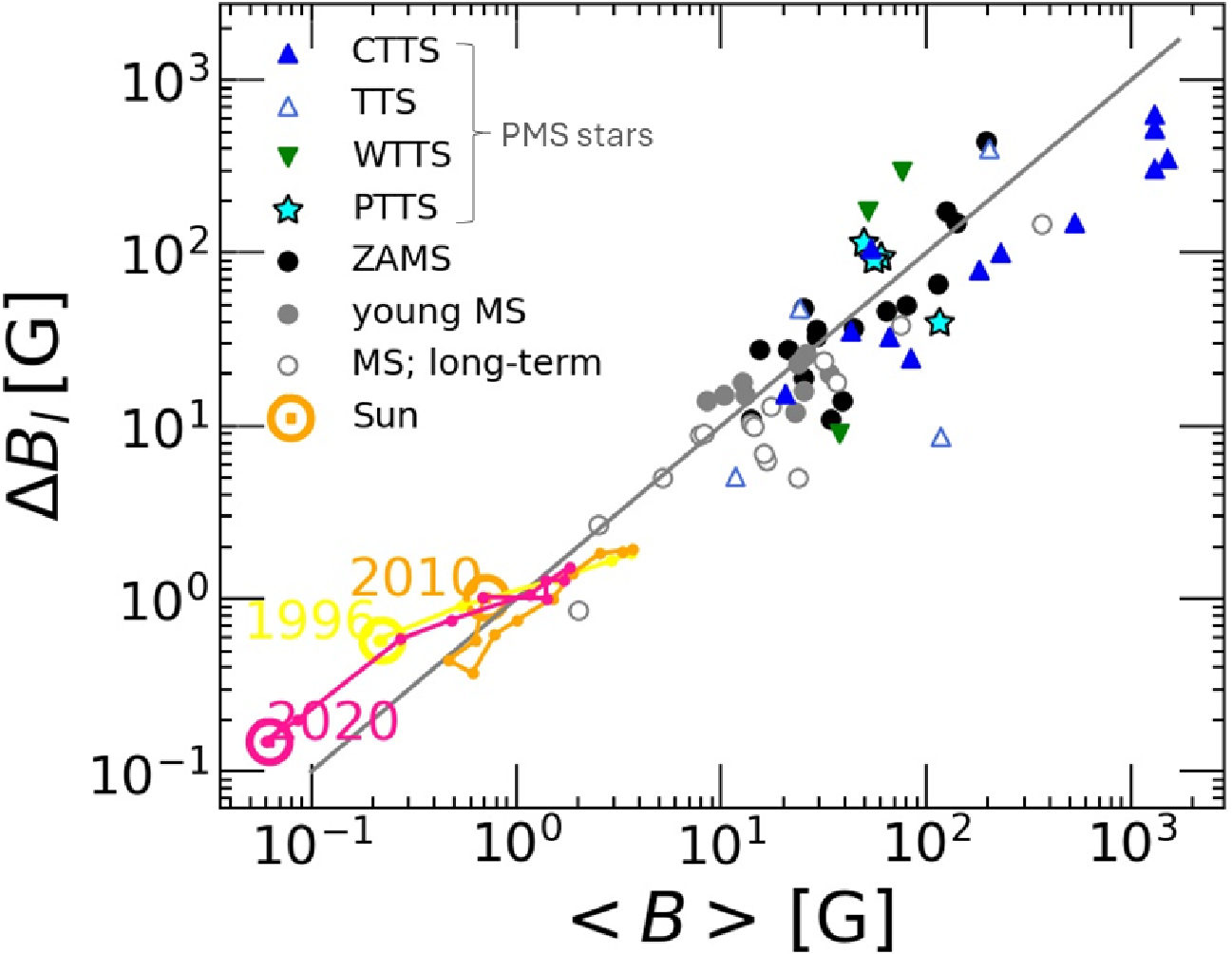}{0.48\textwidth}{(a) All objects \label{figBBa}}
          \fig{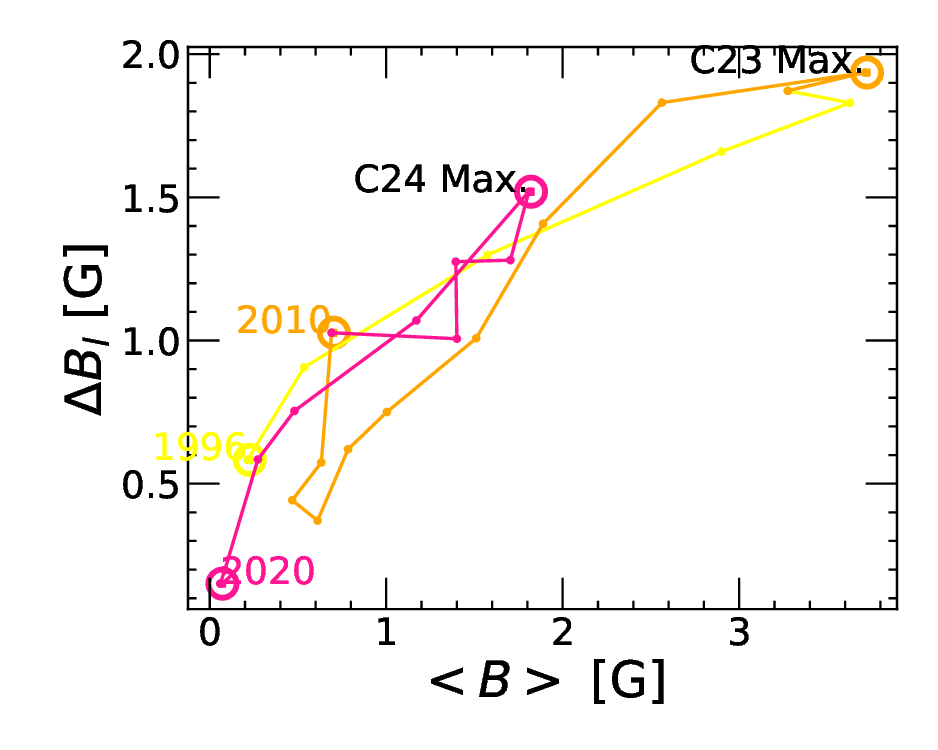}{0.48\textwidth}{(b) Enlarged view of the solar data in \figref{figBBa} \label{figBBs}}
          }
\gridline{\fig{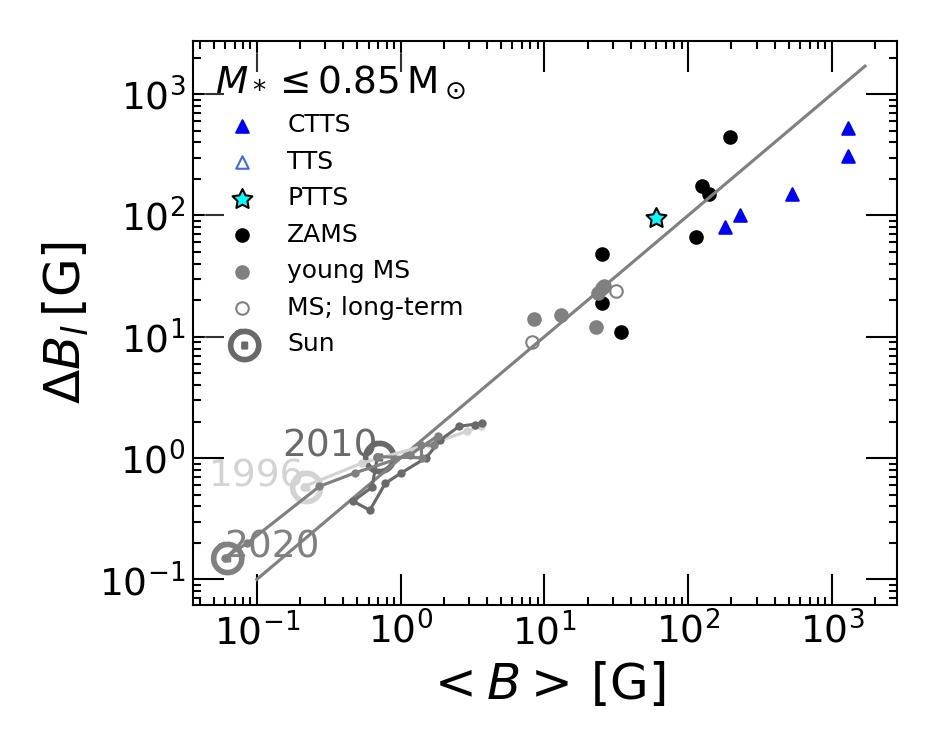}{0.48\textwidth}{(c) $M_* \leq 0.85\ \mathrm{M_\odot}$; $21\ $objects \label{figBBm1}}
          \fig{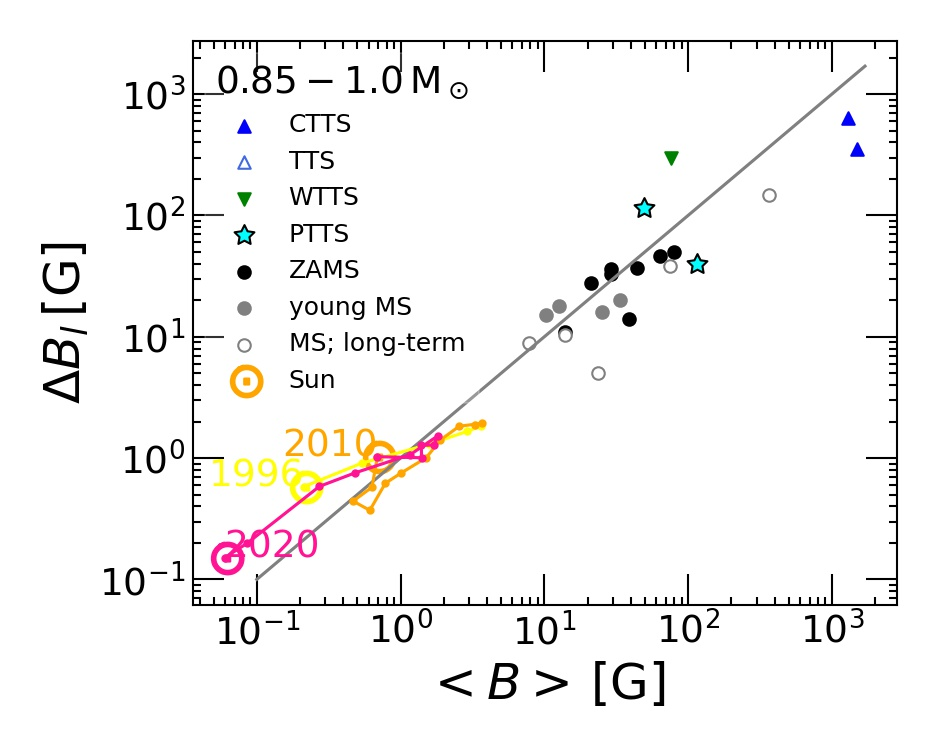}{0.48\textwidth}{(d) $0.85< M_* \leq 1.0\ \mathrm{M_\odot}$; $23\ $objects \label{figBBm2}}
          }
\gridline{\fig{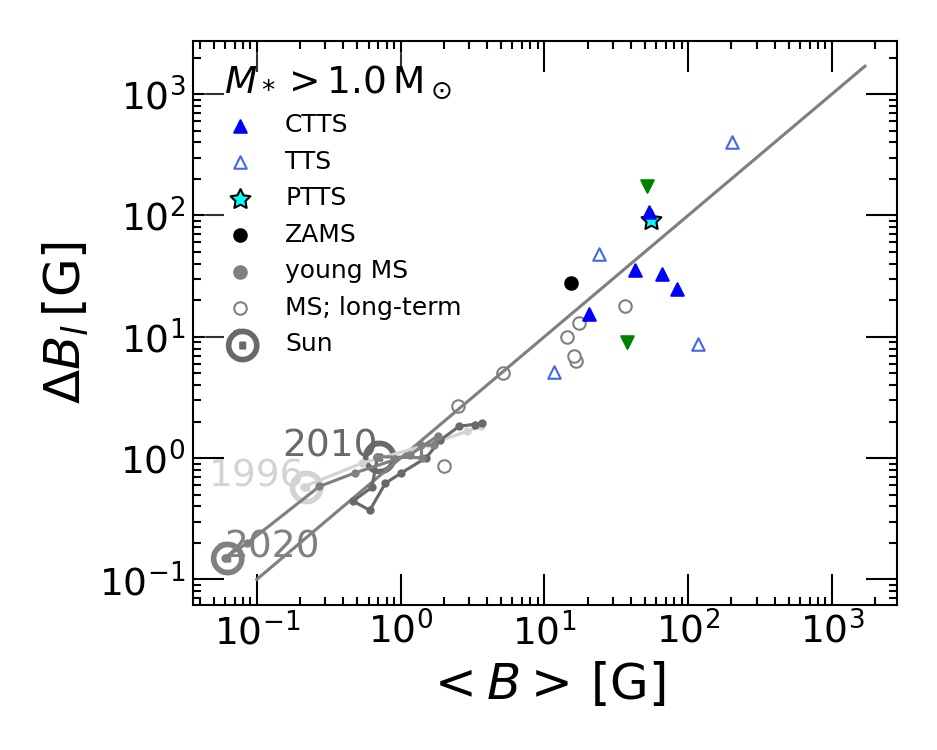}{0.48\textwidth}{(e) $M_* \geq 1.0\ \mathrm{M_\odot}$; $21\ $objects \label{figBBm3}}}
\caption{Relationship between the mean magnetic strength, $<B>$, and their variation, $\Delta B $ of PMS and main-sequence stars. The symbols correspond to \figref{fig:hrdb}. In \Figsref{figBBm1}c, \ref{figBBm2}d, and \ref{figBBm3}e, all objects in \figref{fig:hrdb} are categorized by their mass, $M_*$. The solid line indicates 1:1. }\label{figBB}
\end{figure*}

\Figref{figBB} plots the relationship between the mean magnetic field strengths, $<B>$, and their variations, $\Delta B $, of PMS and main-sequence stars. 
$<B>$ and $\Delta B$ exhibit a positive correlation over three orders of magnitude for various types of low-mass stars. This observation suggests that the mechanism driving the magnetic field is common to PMS stars, main sequence stars, and the Sun. 

The total unsigned flux is often observed in the spatially resolved Sun, which is the sum of the absolute values of the positive and negative magnetic field strengths. The magnetic field data of the Sun must have a higher-order component. For the stellar observations, the total unsigned flux corresponds to the magnetic field strength measured by Zeeman broadening. The Zeeman broadening averaged the magnetic fields in stellar disk, and can be comparable to the disk-averaged solar magnetic field strength. In fact, \citet{ta22} compared the magnetic field data of the Sun and the stellar magnetic flux determined by Zeeman broadening. On the other hand, in this study, the magnetic field strength of our targets was observed by Zeeman-Doppler imaging. In Zeeman-Doppler imaging, the positive and negative polarities cancel each other and the magnetic field strength is underestimated. The effective spatial resolution of the Zeeman-Doppler imaging technique might result in the high-order components of the stellar magnetic fields (i.e., small-scale magnetic structures) of opposite polarities canceling each other out; thus, the total magnetic flux may be underestimated. The details are described in \citet{ta22}. \citet{s19} obtained an empirical power-law relation: 
$<B_{\rm ZB}> = (61\pm17) <B_{\rm ZDI}>^{0.70\pm0.06}$, 
where $<B_{\rm ZB}>$ and $<B_{\rm ZDI}>$ means the magnetic field strength measured by Zeeman broadening and Zeeman-Doppler imaging, respectively. This equation shows that the relationship between $<B_{\rm ZB}>$ and $<B_{\rm ZDI}>$ is almost linear. \Figref{figBB} is then expected to shift to the top right and not distort the current positive correlation relationship. 

The strength of the magnetic field in the solar and stellar spots was only a few kG at most owing to the mechanical balance with the external gas pressure on the stellar surface. 
Assuming that spots are responsible for most of the magnetic field, it is suggested that a large total magnetic field strength indicates a large spot area. 
Yamashita, Itoh, \& Oasa (2025, in prep.) measured the spot coverage of PMS stars using \textit{TESS} photospheric data. The target PMS stars in this study were included in the study by Yamashita, Itoh, \& Oasa (2025, in preparation). A roughly positive correlation was observed between spot coverage and $<B>$. For PMS stars with masses of $0.8 - 1.2\ \mathrm{M_\odot}$ and the Sun, we empirically obtained the following equation: 
\begin{equation}
\log\ ({\rm spot\ coverage}) = -0.155 \times (\log\ <B>)^2 + 1.264 \times (\log\ <B>) -3.246. 
\label{eq:Y24c}
\end{equation}
In this work, the $<B>$ value ranges from $2\ \mathrm{G}$ to $1500\ \mathrm{G}$, which corresponds to the spot coverage of $0.1\%$ to $16\%$, given that \equref{eq:Y24c} is applicable to objects with various masses. 

We did not find a clear difference in $<B>$ and $\Delta B$ between single stars and binaries. All of the three binaries have larger $\Delta B $ than their $<B>$ value. However, they were not plotted quite far from other single stars. 

\citet{i20} assumed that the active regions are inhomogeneously distributed on the surface of an active solar-type star; this assumption was supported by the result in \citet{y22b}. 
The positive correlation between $<B>$ and $\Delta B$ is also shown in \citet{b22} for main-sequence and PMS stars of $T_{\rm eff} = 3200 - 6700\ \mathrm{K}$. Similarly, $R^{\prime}_{\rm HK}$ shows a positive correlation with $\Delta R^{\prime}_{\rm HK}$ in \citet{b22} and \citet{r18} for the Sun and $72$ Sun-like stars. 

One of the key advantages of this study is that the analysis was performed over two solar cycles. The correlated behaviors have been known for the Sun for a very long time, and can be seen in the optical activity for many more cycles (e.g. \citealt{s43}). The data obtained by the two satellites, SOHO and SDO, are homogeneous and have a long observational period. Their data overlapped between the observation periods, and we confirmed the reproducibility during the two cycles. 
In Section \ref{ac}, we discuss the activity cycle of the target objects. Moreover, this study determined the stellar ages and masses. In Sections \ref{ad} and \ref{pm}, we describe the age, while Section \ref{md} covers the mass.

\subsection{Activity cycle} \label{ac}
\citet{r17} analyzed \textit{Kepler} data of 3203 stars and found that the amplitudes of light variation exhibit periodicity with a period of $0.5 - 6\ \mathrm{yr}$, similar to the solar cycle. In \citet{f18}, the dispersion of $<B>$ in main-sequence stars with long-term (multiyear) magnetic variability was attributed to differences in magnetic activity cycles. \Figref{figBB}b presents the $<B>$ and $\Delta B$ of the Sun. $<B>$ and $\Delta B$ vary over the solar cycle. In the solar maximum, both $<B>$ and $\Delta B$ reach their peak values in the cycles, whereas in the solar minimum, they are at their lowest. 
Regarding the other stars, the main-sequence stars in \citet{f18} were observed for six months to six years, whereas the PMS and ZAMS stars were observed for a few days to a few months, an insufficient timeframe for observing the activity cycles. For our targets, the activity cycles were investigated for 34 of the 60 objects in the previous studies (Mount Wilson Observatory monitoring: \citealt{nwv84}, \citealt{o18}, \citealt{s16}, photometry with B, V-band, etc.: \citealt{d17}, \citealt{l16}, \citealt{m02},  \citealt{p10}, \citealt{t11}, X-ray flux variation: \citealt{f20}). The 18 stars have activity cycles of several years, the minimum was 0.2 years, and the maximum was more than 25 years. 
The rotational periods of PMS stars and ZAMS stars in our targets range from $0.4\ \mathrm{days}$ to $13.3\ \mathrm{days}$ (Yamashita, Itoh, \& Oasa 2025, in prep.). Then the main components of the positive correlation are considered to be due to the rotational modulation. 
However, PMS and ZAMS stars still exhibit a positive correlation between $<B>$ and $\Delta B$ and are located at the extensions of the Sun and the main-sequence stars. The variation in solar $<B>$ and $\Delta B$ is considered to result from the activity cycle. As described above, the observational period is not long enough to detect the magnetic activity cycles directly, however, each star has a different period of cycles and the phase in the activity cycle is random in each star when it is observed. Then the variation due to the activity cycle is expected to be added as offset, and the amount of the offset cannot be measured. In \citet{m17}, the relationship between brightness variation caused by starspot and long-term variation (magnetic activity) shows the correlation or anticorrelation. Although it is difficult to separate the offset at this time, it may be possible if long-term observations of V-mag and $<B>$ are conducted for many stars. 
Therefore, we can conclude that not only the activity cycles but also the rotational modulations result in a strong correlation between $<B>$ and $\Delta B$. 

\subsection{Age dependence} \label{ad}
\subsubsection{The relationship between age and magnetic field strength}
\citet{s72} showed that the strength of Ca II HK emission lines, rotation velocity, and lithium abundance decay as $\propto t^{-\frac{1}{2}}$, for the main-sequence stars whose ages are $t = 4\times10^7 - 4.6\times10^{9}\ \mathrm{yr}$ (the Pleiades, Ursa Major, Hyades and the Sun). These decays are considered to lead to spin-down, and as a result, these older slowly rotating stars are unable to generate a strong magnetic field. \citet{h84} also provided the strength of Ca II HK emission lines as a function of age for T Tauri stars. 

In recent years, the Rossby numbers ($N_{\rm R}$ = rotational period / convective turnover time, $\tau_{\rm c}$) are used as the indicator of the magnetic activity. In \citet{f16} and \citet{f18}, the relationship between the Rossby numbers and the mean large-scale magnetic field strength is well established for the ZAMS stars and young main-sequence stars. In \citet{f18}, they found a power-law relationship between $<B>$ and the Rossby numbers: $<B> = (8.4\pm1.8)N_{\rm R}^{-0.89\pm0.13}$ for ZAMS stars and young main-sequence stars with the age of $2.5\times10^8 - 6.5\times10^8\ \mathrm{yr}$. In \citet{f16}, the TTS stars did not follow the trends seen for ZAMS and young main-sequence stars. Their large-scale magnetic field strength is about $10\ $times stronger than that of the ZAMS stars, which extends the trends between the Rossby numbers and $<B>$. However, the rotational periods of TTSs are similar to those of ZAMS stars. They argued that $<B>$ is excessively strong because of the thicker convection zone of TTS and the much longer $\tau_{\rm c}$ compared to older stars.

Within the main-sequence stars, i.e. ZAMS stars, young main-sequence stars, main-sequence stars with long-term (multiyear) magnetic variability, and the Sun, $<B>$ and $\Delta B$ are larger for the younger objects. In each panel of \figref{figBB}, except for panel (b), the ZAMS stars are plotted on the upper right, followed by the young main-sequence stars. The Sun is plotted on the lower left. As main-sequence stars age and their rotational velocities decrease, their ability to generate strong magnetic fields diminishes. 

        	  \begin{figure}[htbp]
        	  \centering
            \includegraphics[width=10cm]{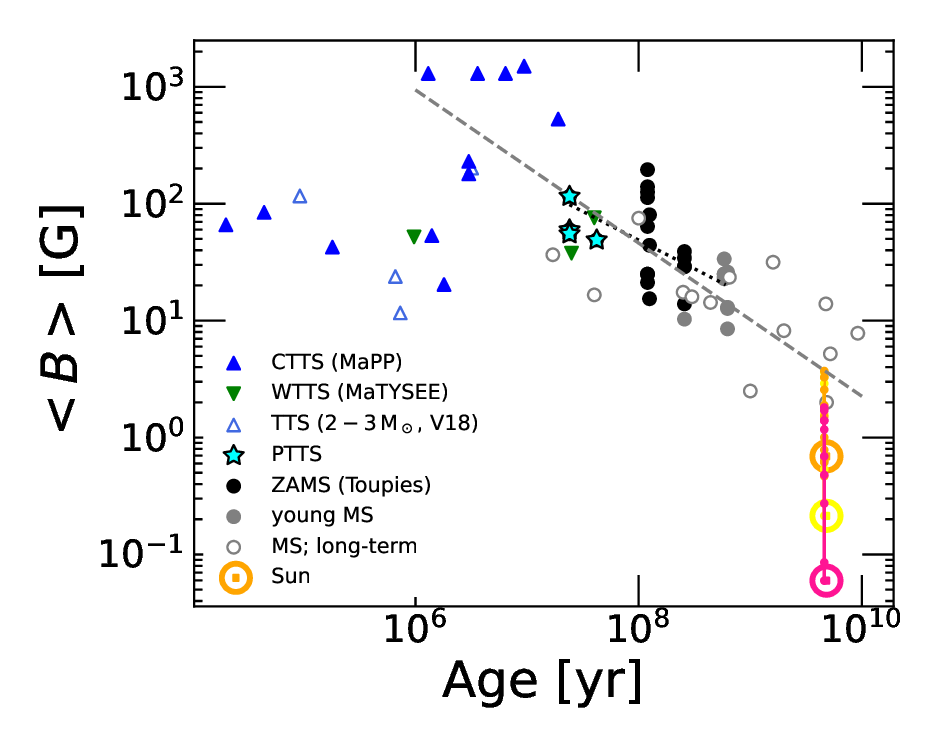}
			\vspace{-0.8cm}
            \caption{ Age dependence of magnetic field strength. The symbols correspond to those in \figref{fig:hrdb}. The dashed and dotted lines represent the decay law presented in \citet{v14} and \citet{f18}, respectively.  } \label{fig:ageB}
        \end{figure}

\Figref{fig:ageB} plots the relationship between the stellar age and the magnetic field strength. We estimated the age of the CTTS, WTTS, and TTS by using the evolutionary models presented by \cite{jk07}. The ages of the PTTSs, ZAMS stars, and young main sequence stars were referred from \citet{f16} and \citet{f18}. The ages of main-sequence stars with long-term (multiyear) magnetic variability were referred from previous studies: \citet{b07}, \citet{b15}, \citet{f10}, Gaia DR3 \citep{g23}, \citet{g00}, \citet{ro16}, and \citet{s00}. However, only II Peg was assumed to be the same as the solar age because its age was not examined. The stellar ages are listed in \tabref{tab:ysoB}. In previous studies, the magnetic field strength decayed with age.
\begin{eqnarray}
<B> & \propto & {\rm age\ \mathrm{[yr]}}^{-0.655\pm0.045}  (\mbox{for}\ 10^6 - 10^{10}\ \mathrm{yr}, \mbox{\citealt{v14}}), \\
<B> & \propto & {\rm (466\pm290)\ age\ \mathrm{[Myr]}}^{-0.49\pm0.12}  (\mbox{for}\ 24 - 600\ \mathrm{Myr}, \mbox{\citealt{f18}}).
\end{eqnarray}
In \figref{fig:ageB}, our targets appear to follow decay laws. Note that all the objects in \citet{f18} are included in our targets.  

\subsubsection{Theoretically predicted spot size}
\citet{mu73} discussed the relationship between the thickness of the convection zone and the size of a sunspot. They predicted that the lower limit of the sunspot size $D_{\rm spot, min}$ increases as the convective layer becomes thicker. Moreover, they considered the depth of the spot, $H_{\rm spot}$, to be equal to the depth of the convection zone, $H_{\rm conv}$, yielding $2.04 < D_{\rm spot, min}/H_{\rm conv} < 6.75$. \citet{y22b} studied the relationship between the convective turnover time, $\tau_{\rm c}$, and the spot coverage $A_{\rm spot} / A_{\rm \frac{1}{2} star}$ for F-, G-, K-type ZAMS stars, and derived $A_{\rm spot} / A_{\rm \frac{1}{2} star} = 10^{-8.7}\cdot \tau_{\rm c}^{1.2} $. 
Then they obtained $D_{\rm spot} \sim 4.3 H_{\rm conv}{}^{1.3}$ by using the relationship between the convective turnover time and $H_{\rm conv}$ derived from \citet{jk07} and \citet{a16}. 

Can we explain the difference in the magnetic field strength by the difference in the size of a single starspot? For a PMS star ($1\ \mathrm{M_\odot}$ and $10^6\ \mathrm{yr}$) and a ZAMS star ($1\ \mathrm{M_\odot}$ and $10^8\ \mathrm{yr}$), we compare the mean magnetic field strength to the spot area: (magnetic field strength of a spot) = (magnetic field strength per unit area) $\times$ (the area of a spot). The calculations are carried out under the following assumptions: 
\begin{itemize}
\item there is only one spot on a stellar surface
\item the magnetic field strength per unit area is constant
\item all the magnetic flux is emitted from starspots, not from faculae or large-scale magnetic fields. 
\end{itemize}
To estimate $D_{\rm spot}$, we estimate $H_{\rm conv}$, which depends on stellar mass and age. $H_{\rm conv}$ can be estimated using the stellar mass and density of convection zone ($M_{\rm conv}$ and $\rho_{\rm conv}$); $H_{\rm conv} = \sqrt[3]{\frac{3M_{\rm conv}}{4\pi\rho_{\rm conv}}}$. We referred the values of $M_{\rm conv}$ and $\rho_{\rm conv}$ from the PMS evolutionary tracks computed by \citet{dm94}.
For example, 
$(M_{\rm conv}, \rho_{\rm conv}) = (1\ \mathrm{M_\odot}, 10^{-0.113}\ \mathrm{g\cdot cm^{-3}})$ for a PMS star of $1\ \mathrm{M_\odot}$ and $10^6\ \mathrm{yr}$. $(M_{\rm conv}, \rho_{\rm conv}) = (0.026\ \mathrm{M_\odot}, 10^{-0.547}\ \mathrm{g\cdot cm^{-3}})$ for a ZAMS star with $1\ \mathrm{M_\odot}$ and $10^8\ \mathrm{yr}$. Thus, 
\begin{eqnarray}
\nonumber
\frac{D_{\rm spot, 1\ \mathrm{M_\odot}, 10^6\ \mathrm{yr}}}{D_{\rm spot, 1\ \mathrm{M_\odot}, 10^8\ \mathrm{yr}}} & \sim & \frac{H_{\rm conv, 1\ \mathrm{M_\odot}, 10^6\ \mathrm{yr}}^{1.3}}{H_{\rm conv, 1\ \mathrm{M_\odot}, 10^8\ \mathrm{yr}}^{1.3}} \\ \nonumber
{} & = & \left( \sqrt[3]{\frac{3M_{\rm conv, 1\ \mathrm{M_\odot}, 10^6\ \mathrm{yr}}}{4\pi\rho_{\rm conv, 1\ \mathrm{M_\odot}, 10^6\ \mathrm{yr}}}} \cdot \left( \sqrt[3]{\frac{3M_{\rm conv, 1\ \mathrm{M_\odot}, 10^8\ \mathrm{yr}}}{4\pi\rho_{\rm conv, 1\ \mathrm{M_\odot}, 10^8\ \mathrm{yr}}}} \right)^{-1} \right)^{1.3} \\ 
{} & = & 3.14. \label{eq:DD}
\end{eqnarray}
We assumed that $D_{\rm spot} \sim 4.3 H_{\rm conv}{}^{1.3}$ is also applied to a PMS star. 

We note that PMS stars with $10^6\ \mathrm{yr}$ are in gravitational collapse, and have larger radius and larger surface area than main-sequence stars. We also note that the magnetic field strength of a star spot is not observable, but the stellar magnetic field strength diluted by the quiet photosphere was observed. Then the spot coverage should be estimated under the consideration of the difference in the surface area. The stellar radius of a solar-mass PMS stars with $10^6\ \mathrm{yr}$ is estimated to be $2.2\ \mathrm{R_\odot}$ with the typical effective temperature and luminosity, ($T_{\rm eff}$, $Lum$) = ($4400\ \mathrm{K}$, $1.6\ \mathrm{L_\odot}$). Considering the difference in the stellar surface area ($= \pi R_*{}^2$), the spot coverage at $10^6\ \mathrm{yr}$ was estimated as 
\begin{equation}
\frac{3.14^2}{2.2^2} = 2.03 \label{eq:DD2}
\end{equation}
times larger than that at $10^8\ \mathrm{yr}$. Then the $<B>$ observed in the CTTSs ($\sim 10^6\ \mathrm{yr}$) with $1\ \mathrm{M_\odot}$ should be $2.03$ times greater than that of the ZAMS stars ($\sim 10^8\ \mathrm{yr}$) with $1\ \mathrm{M_\odot}$. We assume that the magnetic flux of the spots does not vary with the stellar age. However, actually, the average values of observed $<B>$ are $687\pm2\ \mathrm{G}$ and $56\pm2\ \mathrm{G}$ for CTTSs and the ZAMS stars, respectively. The difference in the observed magnetic field strengths at $10^6\ \mathrm{yr}$ and $10^8\ \mathrm{yr}$ was larger than the numerical prediction. 

The results of this study indicate that the difference in magnetic field strength cannot be explained only by the difference in the size of a single starspot and thus the thickness of the convection zone. 
We should deny the assumption that ``there is only one spot on a stellar surface''. 
In \citet{h95}, Doppler images of the WTTS, V410 Tau, indicate that a huge polar spot often comprises several smaller spots.

\subsubsection{Numbers of the spots or spot groups}
Alternatively, in addition to the coverage of a single starspot being limited to twice as much, we hypothesize that the number of starspots on a CTTS is several times that of ZAMS stars. We estimate the number of stellar spots simply from the ratio of the spot area: (mean magnetic field strength) = (magnetic field strength per unit area) $\times$ (the number of the starspots) $\times$ (the area of a spot). The calculations are carried out under the following assumptions: 
\begin{itemize}
\item there is only one spot group on a stellar surface
\item a spot group consists of multiple stellar spots
\item the magnetic field strength per unit area is constant
\item every spots has a radius $D_{\rm spot}$
\item all the magnetic flux is emitted from starspots, not from faculae or large-scale magnetic fields. 
\end{itemize}
For simplicity, we do not consider the case of multiple starspot groups. \citet{ba22} claimed that the lifetimes could be reflective either of the extended existence of a coherent spot group or the emergence of magnetic flux over an extended period in a fairly large area on a star. However, they did not have the spatial resolution to distinguish these. 
Although \citet{ik20} developed a precise method to estimate the number of sunspot groups from TESS and Kepler light curves, it is beyond the scope of this paper to apply this method to dozens of objects. 

The number of spots is estimated in two ways. 
\begin{enumerate}
  \item[A)] Following \equref{eq:DD}, (\ref{eq:DD2}) and the average values of observed $<B>$ are $687\pm2\ \mathrm{G}$ and $56\pm2\ \mathrm{G}$ for CTTSs and the ZAMS stars, the number of starspots is estimated to be six times larger;   
\begin{eqnarray}
  \nonumber
  \frac{4\pi(2.2\ \mathrm{R_\odot})^2}{4\pi(1.0\ \mathrm{R_\odot})^2}\cdot\frac{687\ \mathrm{G}}{56\ \mathrm{G}} & = & \left(\frac{N_{\rm spot, 1\ \mathrm{M_\odot}, 10^6\ \mathrm{yr}} }{N_{\rm spot, 1\ \mathrm{M_\odot}, 10^8\ \mathrm{yr}} }\right) \cdot \left(\frac{\pi D_{\rm spot, 1\ \mathrm{M_\odot}, 10^6\ \mathrm{yr}}{}^2 }{\pi D_{\rm spot, 1\ \mathrm{M_\odot}, 10^8\ \mathrm{yr}}{}^2 }\right) \\
  \therefore \frac{N_{\rm spot, 1\ \mathrm{M_\odot}, 10^6\ \mathrm{yr}} }{N_{\rm spot, 1\ \mathrm{M_\odot}, 10^8\ \mathrm{yr}} } & = & 6.0.
\end{eqnarray}

  \item[B)] The spot coverages are estimated to be $12.4\%$ and $3.1\%$ for CTTSs and the ZAMS stars by substituting their average values of observed $<B>$, $687\ \mathrm{G}$ and $56\ \mathrm{G}$ into \equref{eq:Y24c}. Thus, the number of starspots was estimated to be twice as large: 
\end{enumerate}

\begin{eqnarray}
\nonumber
\mbox{(the stellar surface area)} \cdot \mbox{(the spot coverage)} & = & \mbox{(the number of the starspots)} \cdot \mbox{(the area of a spot)} \\
\nonumber
\frac{4\pi(2.2\ \mathrm{R_\odot})^2}{4\pi(1.0\ \mathrm{R_\odot})^2}\cdot\frac{12.4\%}{3.1\%} & = & \left(\frac{N_{\rm spot, 1\ \mathrm{M_\odot}, 10^6\ \mathrm{yr}} }{N_{\rm spot, 1\ \mathrm{M_\odot}, 10^8\ \mathrm{yr}} }\right) \cdot \left(\frac{\pi D_{\rm spot, 1\ \mathrm{M_\odot}, 10^6\ \mathrm{yr}}{}^2 }{\pi D_{\rm spot, 1\ \mathrm{M_\odot}, 10^8\ \mathrm{yr}}{}^2 }\right) \\
\therefore \frac{N_{\rm spot, 1\ \mathrm{M_\odot}, 10^6\ \mathrm{yr}} }{N_{\rm spot, 1\ \mathrm{M_\odot}, 10^8\ \mathrm{yr}} } & = & 1.96.
\label{eq:honmakaina}
\end{eqnarray}
Here, we claim that the number of spots is two to six times larger on PMS stars than on ZAMS stars. 

The spot coverage of ZAMS stars ($3.1\%$) is similar to the upper limit of the sunspot ($2.7\%$; \citealt{st23}). Their hypothetical simulation demonstrated that the total area of sunspots could be increased to $2.7\%$ of the visible disk by assuming an enhanced lifetime for large sunspots. The largest sunspot coverage over the past 147 years is $1.67\%$ of the visible disk. 
It is possible that the giant starspot in ZAMS may consist of a single spot. Then the lower limit of the number of spots in the CTTS spot group is considered to be four. 

We hypothesize that the large number of starspots in the starspot group may result in the long-lived starspots observed from PMS stars. 
\citet{h95} suggested that this huge spot survived on the stellar surface for approximately $20\ \mathrm{yr}$. 
\citet{bh14} indicated that the larger starspots have longer lifetimes on the Sun and other stars including V410 Tau. 
\citet{c21} investigated optical radial velocity data over a span of $14\ \mathrm{yr}$ for the WTTS, Hubble 4 (V1023 Tau), and found the lifetime of a large spot group to be at least $5.1\ \mathrm{yr}$. 
\citet{ba22} estimated the starspot lifetimes with the Kepler data, and considered that the lifetimes probably refer more to starspot groups than individual spots. 
We can not distinguish whether each sunspot has a long lifespan or appears more frequently. Anyway, it is suggested that new starspots may appear before starspots disappear. As a result, the starspot group has more starspots, and the star spot group is observed as longer apparent lifetime. 


\subsection{PMS stars} \label{pm}
The PMS stars (CTTS, TTS, WTTS, and PTTS) exhibited the strongest magnetic fields of all the objects.  CTTSs ($\approx 10^6\ \mathrm{yr}$) have particularly large $<B>$ and $\Delta B$. However, their $\Delta B$ values fall below the linear proportionality line extrapolated from aged stars. This can be explained in three ways. 
The first possibility is that the massive spot is located near the pole; thus, the spot is not entirely hidden by stellar rotation. Using Zeeman--Doppler imaging, previous studies found that some WTTS and young main-sequence stars have large spots at high latitudes (e.g., V410 Tau and EK Dra in \citealt{rs96}, \citealt{sr98}, and others). \citet{y15} demonstrated dynamo simulation with a strong dipolar magnetic field aligned with the rotation axis for objects with thick convective zones (e.g. PMS stars), reporting that this effect resulted in the large northern spot. 
Second, strong magnetic fields are distributed throughout the stellar surface, which reduces the variation in the magnetic field strength resulting from stellar rotational modulation. The magnetic filling factor of the photosphere tends to increase with the stellar rotation or magnetic activity level (e.g., \citealt{r09} for M dwarfs). In fact, AA Tau, one of our samples with a magnetic field strength greater than $1000\ \mathrm{G}$, is almost completely filled with the magnetic fields of negative magnetic components in the Zeeman--Doppler imaging \citep{d10}. 
Third, $\Delta B$ may be underestimated because the PMS stars are faint, a longer exposure time is required, and the number of observations (i.e., sampling number) is limited. \citet{b22} also found that $\Delta B$ are smaller for objects with fewer observations. 

\subsection{Mass dependence} \label{md}
In \Figsref{figBB}c, \ref{figBB}d, and \ref{figBB}e, objects with a higher mass show smaller $<B>$ and $\Delta B$. The average values of $<B>$ are $66\pm4\ \mathrm{G}$, $49\pm4\ \mathrm{G}$, and $25\pm3\ \mathrm{G}$ for objects with $M_* \leq 0.85\ \mathrm{M_\odot}$, $0.85< M_* \leq 1.0\ \mathrm{M_\odot}$, and $M_* \geq 1.0\ \mathrm{M_\odot}$, respectively. 
These $<B>$ values can be converted to the spot coverage (the total area of starspots against the visible area of the stellar hemisphere) of $3.5\%$, $2.8\%$, and $1.7\%$ using \equref{eq:Y24c}. This is because objects with higher masses are expected to have thinner convection zones. In this subsection, we calculate the ratio of the spot sizes in the two aforementioned cases in the same manner as in \equref{eq:DD}.

\subsubsection{$0.5\ \mathrm{M_\odot}$ and $1\ \mathrm{M_\odot}$ stars at ZAMS }

\begin{enumerate}
  \item[A)] The difference between the spot size of a ZAMS star with $0.5\ \mathrm{M_\odot}$ and that with $1\ \mathrm{M_\odot}$ was calculated as follows:
\begin{eqnarray}
\nonumber
\frac{D_{\rm spot, 0.5\ \mathrm{M_\odot}, 10^8\ \mathrm{yr}}}{D_{\rm spot, 1\ \mathrm{M_\odot}, 10^8\ \mathrm{yr}}} & \sim & \frac{H_{\rm conv, 0.5\ \mathrm{M_\odot}, 10^8\ \mathrm{yr}}^{1.3}}{H_{\rm conv, 1\ \mathrm{M_\odot}, 10^8\ \mathrm{yr}}^{1.3}} \\ 
{} & = & 0.43. \label{eq:DD3}
\end{eqnarray}
In order to estimate $H_{\rm conv} = \sqrt[3]{\frac{3M_{\rm conv}}{4\pi\rho_{\rm conv}}}$, we referred $(M_{\rm conv}, \rho_{\rm conv}) = (0.137\ \mathrm{M_\odot}, 10^{1.025}\ \mathrm{g\cdot cm^{-3}})$ and $(M_{\rm conv}, \rho_{\rm conv}) = (0.026\ \mathrm{M_\odot}, 10^{-0.547}\ \mathrm{g\cdot cm^{-3}})$ for a ZAMS star with $0.5\ \mathrm{M_\odot}$ and that with $1\ \mathrm{M_\odot}$, respectively.
The stellar radius of a ZAMS stars with $1\ \mathrm{M_\odot}$ is $1\ \mathrm{R_\odot}$, and that of a ZAMS stars with $0.5\ \mathrm{M_\odot}$ is estimated to be $0.43\ \mathrm{R_\odot}$ with the typical effective temperature and luminosity in \citet{dm94}, ($T_{\rm eff}$, $Lum$) = ($3917\ \mathrm{K}$, $0.04\ \mathrm{L_\odot}$). Thus, the spot coverage of a ZAMS star with $0.5\ \mathrm{M_\odot}$ is equal to that of a ZAMS star with $1\ \mathrm{M_\odot}$: 
\begin{equation}
\frac{0.43^2}{(0.43/1.00)^2} = 1.00. \label{eq:a9}
\end{equation}
In contrast, the average values of $<B>$ for the ZAMS stars with $M_* \leq 0.85\ \mathrm{M_\odot}$ and $0.85< M_* \leq 1.0\ \mathrm{M_\odot}$ are $79\pm2\ \mathrm{G}$ and $46\pm2\ \mathrm{G}$, respectively. Thus, the ratio of the magnetic field strength was $79\ \mathrm{G} / 46\ \mathrm{G} = 1.72$. By using \equref{eq:DD3} and (\ref{eq:a9}), the number of starspots in ZAMS stars with $0.5\ \mathrm{M_\odot}$ may be $1.72/1.00 = 1.72$ times greater than that in ZAMS stars with $1\ \mathrm{M_\odot}$: 
\begin{eqnarray}
\nonumber
\frac{4\pi(0.43\ \mathrm{R_\odot})^2}{4\pi(1.0\ \mathrm{R_\odot})^2}\cdot\frac{79\ \mathrm{G}}{46\ \mathrm{G}} & = & \left(\frac{N_{\rm spot, 0.5\ \mathrm{M_\odot}, 10^8\ \mathrm{yr}} }{N_{\rm spot, 1\ \mathrm{M_\odot}, 10^8\ \mathrm{yr}} }\right) \cdot \left(\frac{\pi D_{\rm spot, 0.5\ \mathrm{M_\odot}, 10^8\ \mathrm{yr}}{}^2 }{\pi D_{\rm spot, 1\ \mathrm{M_\odot}, 10^8\ \mathrm{yr}}{}^2 }\right) \\
\therefore \frac{N_{\rm spot, 0.5\ \mathrm{M_\odot}, 10^8\ \mathrm{yr}} }{N_{\rm spot, 1\ \mathrm{M_\odot}, 10^8\ \mathrm{yr}} } & = & 1.72. 
\end{eqnarray}

  \item[B)] With the average values of $<B>$ ($79\ \mathrm{G}$ and $46\ \mathrm{G}$) and \equref{eq:Y24c}, the spot coverages are estimated to be $4.0\%$ and $2.7\%$ for a ZAMS star with $0.5\ \mathrm{M_\odot}$ and that with $1\ \mathrm{M_\odot}$, respectively. Then similar to the discussion in \equref{eq:honmakaina}, the number of the starspots is estimated with the stellar surface area, the spot coverage estimated above ($4.0\%$ and $2.7\%$), and the spot size in \equref{eq:DD3}: 
\begin{eqnarray}
\nonumber
\frac{4\pi(0.43\ \mathrm{R_\odot})^2}{4\pi(1.0\ \mathrm{R_\odot})^2}\cdot\frac{4.0\%}{2.7\%} & = & \left(\frac{N_{\rm spot, 0.5\ \mathrm{M_\odot}, 10^8\ \mathrm{yr}} }{N_{\rm spot, 1\ \mathrm{M_\odot}, 10^8\ \mathrm{yr}} }\right) \cdot \left(\frac{\pi D_{\rm spot, 0.5\ \mathrm{M_\odot}, 10^8\ \mathrm{yr}}{}^2 }{\pi D_{\rm spot, 1\ \mathrm{M_\odot}, 10^8\ \mathrm{yr}}{}^2 }\right) \\
\therefore \frac{N_{\rm spot, 0.5\ \mathrm{M_\odot}, 10^8\ \mathrm{yr}} }{N_{\rm spot, 1\ \mathrm{M_\odot}, 10^8\ \mathrm{yr}} } & = & 1.48. 
\label{eq:honmakaina2}
\end{eqnarray}
The number of starspots was calculated to be $1.48$ times larger, assuming that \equref{eq:Y24c} is applicable to objects with various masses. 
\end{enumerate}

\subsubsection{$0.5\ \mathrm{M_\odot}$ and $2\ \mathrm{M_\odot}$ stars at $10^6\ \mathrm{yr}$}

\begin{enumerate}
  \item[A)] The radius ratio between the single spot of a PMS star with $0.5\ \mathrm{M_\odot}$ and that with $2\ \mathrm{M_\odot}$ was calculated as follows:
\begin{eqnarray}
\nonumber
\frac{D_{\rm spot, 0.5\ \mathrm{M_\odot}, 10^6\ \mathrm{yr}}}{D_{\rm spot, 2\ \mathrm{M_\odot}, 10^6\ \mathrm{yr}}} & \sim & \frac{H_{\rm conv, 0.5\ \mathrm{M_\odot}, 10^6\ \mathrm{yr}}^{1.3}}{H_{\rm conv, 2\ \mathrm{M_\odot}, 10^6\ \mathrm{yr}}^{1.3}} \\ 
{} & = & 0.40. \label{eq:DD4}
\end{eqnarray}
In order to estimate $H_{\rm conv} = \sqrt[3]{\frac{3M_{\rm conv}}{4\pi\rho_{\rm conv}}}$, we referred $(M_{\rm conv}, \rho_{\rm conv}) = (0.500\ \mathrm{M_\odot}, 10^{-0.072}\ \mathrm{g\cdot cm^{-3}})$ and $(M_{\rm conv}, \rho_{\rm conv}) = (1.359\ \mathrm{M_\odot}, 10^{-0.555}\ \mathrm{g\cdot cm^{-3}})$ for a PMS star with $0.5\ \mathrm{M_\odot}$ and that with $2\ \mathrm{M_\odot}$, respectively.
When the age of PMS stars are $10^6\ \mathrm{yr}$, the radius of the PMS star with $0.5\ \mathrm{M_\odot}$ and $2\ \mathrm{M_\odot}$ are estimated to be $1.72\ \mathrm{R_\odot}$ and $3.21\ \mathrm{R_\odot}$ by substituting the typical effective temperature and luminosity in \citet{dm94}, ($T_{\rm eff}$, $Lum$) = ($3837\ \mathrm{K}$, $0.56\ \mathrm{L_\odot}$) and ($T_{\rm eff}$, $Lum$) = ($5105\ \mathrm{K}$, $6.31\ \mathrm{L_\odot}$) into the Stefan-Boltzmann law. The spot coverage of a PMS star with $0.5\ \mathrm{M_\odot}$ can be estimated to be 0.56 times larger than that at $2\ \mathrm{M_\odot}$: 
\begin{equation}
\frac{0.40^2}{(1.72/3.21)^2} = 0.56. \label{eq:a11}
\end{equation}
In contrast, the average values of $<B>$ are $517\pm3\ \mathrm{G}$ and $48\pm2\ \mathrm{G}$ for the CTTSs with $M_* \leq 0.85\ \mathrm{M_\odot}$ and $M_* \geq 1.0\ \mathrm{M_\odot}$, respectively. Thus, the magnetic field strength ratio is $517\ \mathrm{G} / 48\ \mathrm{G} = 10.77$. The number of starspots on PMS stars with $0.5\ \mathrm{M_\odot}$ may be $10.77/0.56 = 19.33$ times more than that on PMS stars with $2\ \mathrm{M_\odot}$. 
\begin{eqnarray}
\nonumber
\frac{4\pi(1.72\ \mathrm{R_\odot})^2}{4\pi(3.21\ \mathrm{R_\odot})^2}\cdot\frac{517\ \mathrm{G}}{48\ \mathrm{G}} & = & \left(\frac{N_{\rm spot, 0.5\ \mathrm{M_\odot}, 10^6\ \mathrm{yr}} }{N_{\rm spot, 2\ \mathrm{M_\odot}, 10^6\ \mathrm{yr}} }\right) \cdot \left(\frac{\pi D_{\rm spot, 0.5\ \mathrm{M_\odot}, 10^6\ \mathrm{yr}}{}^2 }{\pi D_{\rm spot, 2\ \mathrm{M_\odot}, 10^6\ \mathrm{yr}}{}^2 }\right) \\
\therefore \frac{N_{\rm spot, 0.5\ \mathrm{M_\odot}, 10^6\ \mathrm{yr}} }{N_{\rm spot, 2\ \mathrm{M_\odot}, 10^6\ \mathrm{yr}} } & = & 19.33. 
\end{eqnarray}

  \item[B)] The average values of $<B>$, $517\ \mathrm{G}$ and $48\ \mathrm{G}$ corresponds to the spot coverage of $11\%$ and $2.8\%$ with \equref{eq:Y24c}, respectively.  Similar to the discussion in \equref{eq:honmakaina}, the number of the starspots is estimated with the stellar surface area, the estimated spot coverage, and the spot size in \equref{eq:DD4}: 
\begin{eqnarray}
\nonumber
\frac{4\pi(1.72\ \mathrm{R_\odot})^2}{4\pi(3.21\ \mathrm{R_\odot})^2}\cdot\frac{11\%}{2.8\%} & = & \left(\frac{N_{\rm spot, 0.5\ \mathrm{M_\odot}, 10^6\ \mathrm{yr}} }{N_{\rm spot, 2\ \mathrm{M_\odot}, 10^6\ \mathrm{yr}} }\right) \cdot \left(\frac{\pi D_{\rm spot, 0.5\ \mathrm{M_\odot}, 10^6\ \mathrm{yr}}{}^2 }{\pi D_{\rm spot, 2\ \mathrm{M_\odot}, 10^6\ \mathrm{yr}}{}^2 }\right) \\
\therefore \frac{N_{\rm spot, 0.5\ \mathrm{M_\odot}, 10^6\ \mathrm{yr}} }{N_{\rm spot, 2\ \mathrm{M_\odot}, 10^6\ \mathrm{yr}} } & = & 7.05. 
\end{eqnarray}
The number of starspots on a $0.5\ \mathrm{M_\odot}$ PMS star was calculated to be $7.05$ times more than a $2\ \mathrm{M_\odot}$ PMS star, assuming that \equref{eq:Y24c} is applicable to objects with various masses. 
\end{enumerate}

These results may be highly indeterminate. Since there are only a few observations of starspot lifetimes, further observations are needed to confirm the number of starspots estimated in this study. 

\section{Conclusion}

This study investigated the relationship between the disk-averaged magnetic field strength (in the line-of-sight direction), $<B>$, and the variation of the magnetic field strength, $\Delta B = B_{\rm max} - B_{\rm min}$ for the Sun, main-sequence stars, and TTSs. The targets are limited to stars where magnetic field variations have been observed in previous studies because such observations are few and rare. We mainly referred to the previous studies examined systematically $<B>$ and $\Delta B$. The targets are $28\ $PMS stars, $17\ $ZAMS stars, $10\ $young main-sequence stars, $15\ $main-sequence stars with long-term (multi-year) magnetic variability, and the Sun. 
For the Sun, we analyzed the time series of the averaged unsigned magnetic strength of the Sun observed with SOHO/MDI and SDO/HMI from 1996 to 2019. The data from the two satellites cover two solar cycles, solar cycles 23 and 24. 

The following key conclusions were drawn based on the analysis. 
\begin{enumerate}
\item[(1)] $<B>$ and $\Delta B$ exhibit a positive correlation over three orders of magnitude, suggesting that the mechanism driving the magnetic field is common to the Sun, main-sequence stars, and TTSs. The observed positive correlation further implies that stars with larger spot sizes and numbers exhibit larger variation amplitudes due to rotational modulations. 
We checked binaries or triplets listed in Gaia DR3 \citep{g23}. However, we did not find a clear difference in $<B>$ and $\Delta B$ between single stars and binaries. Their $<B>$ and $\Delta B$ values are not quite larger or smaller than other single stars. 
We discussed $<B>$ and $\Delta B$ in terms of the activity cycles, age dependence, and mass dependence on the starspots.

\item[(2)] Activity cycles: In the case of the Sun, both $<B>$ and $\Delta B$ tend to be larger during solar maximum compared to that during solar minimum. The target stars except for main-sequence stars with long-term (multi-year) magnetic variability were observed in $4\ \mathrm{days}$ to $6\ \mathrm{years}$. The timescale of observations for main-sequence stars with long-term (multi-year) magnetic variability is half a year to $9\ \mathrm{years}$. For our targets, the activity cycles were investigated for 34 of the 60 objects in the previous studies. The 18 stars have activity cycles of several years, the minimum was 0.2 years, and the maximum was more than 25 years. The rotational periods of PMS stars and ZAMS stars in our targets range from $0.4\ \mathrm{days}$ to $13.3\ \mathrm{days}$ (Yamashita, Itoh, \& Oasa 2025, in prep.). Then the main components of the positive correlation are considered to be due to the rotational modulation. The observational period is not long enough to detect the magnetic activity cycles directly, however, each star has a different period of cycles and it is random which activity period is observed. Therefore, we can conclude that not only the activity cycles but also the rotational modulations result in a strong correlation between $<B>$ and $\Delta B$.

\item[(3)] Age dependence: The younger objects (e.g. CTTS) have larger $<B>$ and $\Delta B$, and older objects (e.g. main-sequence stars and the Sun) have smaller $<B>$ and $\Delta B$. $<B>$ of our targets decrease with their age, and approximately followed the decay laws presented in \citet{v14} and \citet{f18}. 

\citet{mu73} discussed the relationship between the thickness of the convection zone and the size of a sunspot. \citet{y22b}  obtained $D_{\rm spot} \sim 4.3 H_{\rm conv}{}^{1.3}$ by using the relationship between the convective turnover time and $H_{\rm conv}$ derived from the previous study. In this study, we estimated $H_{\rm conv} = \sqrt[3]{\frac{3M_{\rm conv}}{4\pi\rho_{\rm conv}}}$ using the stellar mass and density of convection zone ($M_{\rm conv}$ and $\rho_{\rm conv}$) from the PMS evolutionary tracks computed by \citet{dm94}. Then we predicted that the spot coverage of a CTTS ($1\ \mathrm{M_\odot}$ and $10^6\ \mathrm{yr}$) is twice larger than that of a ZAMS star ($1\ \mathrm{M_\odot}$ and $10^8\ \mathrm{yr}$) if they have single starspot. This result is not consistent with their average values of observed $<B>$: CTTSs have twelve times stronger magnetic fields than ZAMS stars. 

Then we hypothesize that the number of starspots on a CTTS is several times that of ZAMS stars. We estimate the number of stellar spots, which are two to six times larger on PMS stars than on ZAMS stars. The large number of starspots in the starspot group is considered to contribute to the long-lived starspots observed from PMS stars. In \citet{ma17} and \citet{na19}, starspot lifetimes of solar-type main-sequence stars have been observed to range from tens to hundreds of days. It should be noted that the duration and frequency of observations are different, but starspots with longer lifetimes than main-sequence stars have been detected from PMS stars (e.g. \citealt{h95}). 
It is suggested that a possible scenario in which new spots are consistently born, before old spots disappear. As a result, the starspot group has more starspots, and the star spot group is observed as longer apparent lifetime. 

Furthermore, $\Delta B$ values of CTTSs fall below the linear proportionality line extrapolated from aged stars. The first possibility is that the massive spot is located near the pole. Second, strong magnetic fields are distributed throughout the stellar surface, which reduces the variation in the magnetic field strength resulting from stellar rotational modulation. Third, $\Delta B$ may be underestimated because the PMS stars are faint, a longer exposure time is required, and the number of observations (i.e., sampling number) is limited. One or all of these three may be responsible for the $\Delta B$ decrease since previous studies support any of the hypotheses.

\item[(4)] Mass dependence: With $H_{\rm conv}$ and $<B>$, we estimated the number of stellar spots is about two times larger on a ZAMS stars with $0.5\ \mathrm{M_\odot}$ than a ZAMS stars with $1\ \mathrm{M_\odot}$.This result is supported by the decay laws between effective temperature and starspot lifetime \citep{g17}. 
The age differences described above are considered to be equally or more effective in increasing the number of starspots rather than differences in stellar mass. 

In the same way, we estimated the number of starspots on a PMS star at $10^6\ \mathrm{yr}$, and compared $0.5\ \mathrm{M_\odot}$ and $2\ \mathrm{M_\odot}$. The number of starspots on a $0.5\ \mathrm{M_\odot}$ PMS star was calculated to be seven or nineteen times more than a $2\ \mathrm{M_\odot}$ PMS star. These results may be highly indeterminate. Since there are only a few observations of starspot lifetimes, further observations are needed to confirm the number of starspots estimated in this study. 
\end{enumerate}

\begin{acknowledgments}
We thank Prof. Kiyoshi Ichimoto for these discussions. 
This work would not have been possible without financial support from JSPS KAKENHI grant number 23KJ1855. M.Y. was also supported by the JSPS Research Fellows (DC2 and PD). Y. I. was supported by JSPS KAKENHI grant number 17K05390. S.T. was supported by JSPS KAKENHI under grant numbers JP20KK0072, JP21H01124, and JP21H04492. 
\end{acknowledgments}

%

\vspace{5mm}
\facilities{\textit{Transiting Exoplanet Survey Satellite} (\textit{TESS}, \citealt{ri15})}




\bibliography{sample631}{}
\bibliographystyle{aasjournal}



\end{CJK*}
\end{document}